\documentclass[conference]{IEEEtran}
\IEEEoverridecommandlockouts
\usepackage{cite}
\usepackage{amsmath,amssymb,amsfonts}
\usepackage{graphicx}
\usepackage{textcomp}
\usepackage{xcolor}
\usepackage{mathtools} %
\usepackage{todonotes} %
\usepackage[caption=false]{subfig} %
\usepackage[mathscr]{euscript} 
\usepackage{url} %
\usepackage{hyperref}
\usepackage{bm}
\usepackage{algpseudocode} 
\usepackage{algorithm}
\usepackage{xifthen}
\usepackage[export]{adjustbox}

\newcommand{\mycomment}[1]{}
\NewDocumentCommand{\vect}{ O{} O{} m }{\mathbf{#3}\ifthenelse{\isempty{#1}}{}{^{(#1)}}\ifthenelse{\isempty{#2}}{}{_{#2}}}

\NewDocumentCommand{\mat}{ O{} O{} m }{\mathbf{#3}\ifthenelse{\isempty{#1}}{}{^{(#1)}}\ifthenelse{\isempty{#2}}{}{_{#2}}}

\NewDocumentCommand{\ten}{ O{} O{} m }{\pmb{\mathscr{#3}}\ifthenelse{\isempty{#1}}{}{^{(#1)}}\ifthenelse{\isempty{#2}}{}{_{#2}}}

\def\BibTeX{{\rm B\kern-.05em{\sc i\kern-.025em b}\kern-.08em
    T\kern-.1667em\lower.7ex\hbox{E}\kern-.125emX}}
\begin{document}

\title{MalwareDNA: Simultaneous Classification of Malware, Malware Families, and Novel Malware
}



\author{\IEEEauthorblockN{1\textsuperscript{st} Maksim E. Eren}
\IEEEauthorblockA{\textit{Advanced Research in Cyber Systems} \\
\textit{Los Alamos National Laboratory}\\
Los Alamos, USA \\
maksim@lanl.gov}

\and
\IEEEauthorblockN{2\textsuperscript{nd} Manish Bhattarai}
\IEEEauthorblockN{3\textsuperscript{rd} Kim Rasmussen}
\IEEEauthorblockN{4\textsuperscript{th} Boian S. Alexandrov}
\IEEEauthorblockA{\textit{Theoretical Division} \\
\textit{Los Alamos National Laboratory}\\
Los Alamos, USA}

\and
\IEEEauthorblockN{5\textsuperscript{th} Charles Nicholas}
\IEEEauthorblockA{\textit{CSEE} \\
\textit{University of Maryland, Baltimore County}\\
Maryland, USA}

}


\maketitle
\vspace{-40em}

\begin{abstract}
Malware is one of the most dangerous and costly cyber threats to national security and a crucial factor in modern cyber-space. However, the adoption of machine learning (ML) based solutions against malware threats has been relatively slow. Shortcomings in the existing ML approaches are likely contributing to this problem. The majority of current ML approaches ignore real-world challenges such as the detection of novel malware. In addition, proposed ML approaches are often designed either for malware/benign-ware classification or malware family classification. Here we introduce and showcase preliminary capabilities of a new method that can perform precise identification of novel malware families, while also unifying the capability for malware/benign-ware classification and malware family classification into a single framework.
\end{abstract}


\section{Introduction}
Approximately half a million new malware are reported daily, which drives the increased utilization of Machine Learning (ML) based automated security systems to combat malware \cite{avtest_2021}. Several ML solutions have previously been introduced for distinct tasks of malware detection and malware family classification. The objective of malware detection is to identify a given file as benign or malicious. In contrast to malware detection, malware family classification assumes that any given sample is already known to be malicious, and we want to know which family it belongs to \cite{Raff2020ASO}. Existing solutions often use separate ML systems, where one system may be used for detecting malware, and another system is then used to classify the detected malware into a given family. A system that can unify these tasks would have operational benefits such as reducing the complexity of maintaining separate systems.

In addition, despite its benefits, the adoption of ML-based solutions against malware threats has been relatively slow due to shortcomings in these systems \cite{Raff2020ASO}. The majority of the past two decades of research on malware family classification, has not sufficiently accounted for core evaluation criteria including the ability to identify new malware \cite{nguyen2021leveraging, Raff2020ASO}. New malware samples are created regularly by threat actors, which create new versions of already existing malware with identical functionality \cite{Raff2020ASO}. Malware analysts regularly go through large quantities of malware samples to understand whether a new malware specimen belongs to a previously known malware family. Classifying a new malware sample into a family or identifying it as \textit{novel} can reduce the number of files analysts need to examine, and aid in understanding the behavior of the malware; this in turn helps estimating the severity of the threat and developing mitigation strategies \cite{Raff2020ASO}. At the same time, semi-supervised learning in the malware classification field has not been widely explored despite its superior generalization to new data as compared to supervised systems \cite{Raff2020ASO}. With the ever-growing quantity of malware and their complexities there is an urgent need to improve existing solutions and their operational architectures to drive the increased adaption of ML solutions.

Here we introduce a new semi-supervised method, named \textit{MalwareDNA}, that unifies the capability of malware detection and malware family classification into a single framework, while also addressing the shortcomings of novel malware family identification. In this way, MalwareDNA can classify known malware families and separate them from benign-ware, as well as identify new types of families, \textit{all at the same time}. Our method uses hierarchical non-negative matrix factorization (NMF) with automatic model determination \cite{SmartTensors}, which enables data modeling with high specificity and accuracy, to build an archive of latent signatures (identifiers) of malware and benign-ware. These signatures are then be used for precise real-time downstream detection of malware and classification of malware families. Our method also includes a fast optimization method to perform real-time identification of unseen signatures (or novel malware families) by implementation of the \textit{reject-option} method \cite{ding2020revisiting}. To the best of our knowledge, we are the first to introduce a framework that combines malware detection, malware family classification, and novel malware family identification capabilities into a single system.

\section{Relevant Work}

As part of the semi-supervised scheme, our method leverages clustering and similarity scores for categorization of novel samples.  A number of previous works have also used clustering approaches, where the ensemble of clustering algorithms with distinct characteristics has been shown to yield accurate results for malware classification \cite{10.1145/1835804.1835820, 8029425}. Likewise, similarity metrics to extract embeddings (distance-based feature vectors) has also shown to be a successful technique for malware analysis \cite{10.1145/2487575.2488219}. These methods, however, only focus on malware/benign-ware detection or malware family classification, and do not posses the ability to identify novel families.

Several works did consider benign-ware as a class among the classes of malware families \cite{loi2021towards, huang2016mtnet}. This allowed these methods to separate benign-ware from malware and also classify malware families simultaneously. At the same time, these prior works attempted to detect rare specimens by grouping multiple families into a single \textit{"others"} class. The most realistic malware family classification work was done by Huang et al. which targeted 100 classes where two of the classes include one for benign samples and \textit{"others"} \cite{huang2016mtnet}. While this approach introduced an ability to detect rare specimens by the "others" class, it yields poor generalization to new or never before seen specimens as was also pointed out by Loi et al \cite{loi2021towards}. Loi et al. reports that their false positives are heavily represented by the families collected within the \textit{"others"} class due to the supervised method's inability to learn the patterns of these families from a small number of specimens. Conversely, our method does not require training with rare specimens, instead it posses the abstaining prediction ability (the \textit{reject-option}). This allows our method to uniquely combine the abilities of malware detection and malware family classification, as well as novel malware family identification.

\section{Method}
\begin{figure}[htb]
  \centerline{\includegraphics[width=0.75\linewidth]{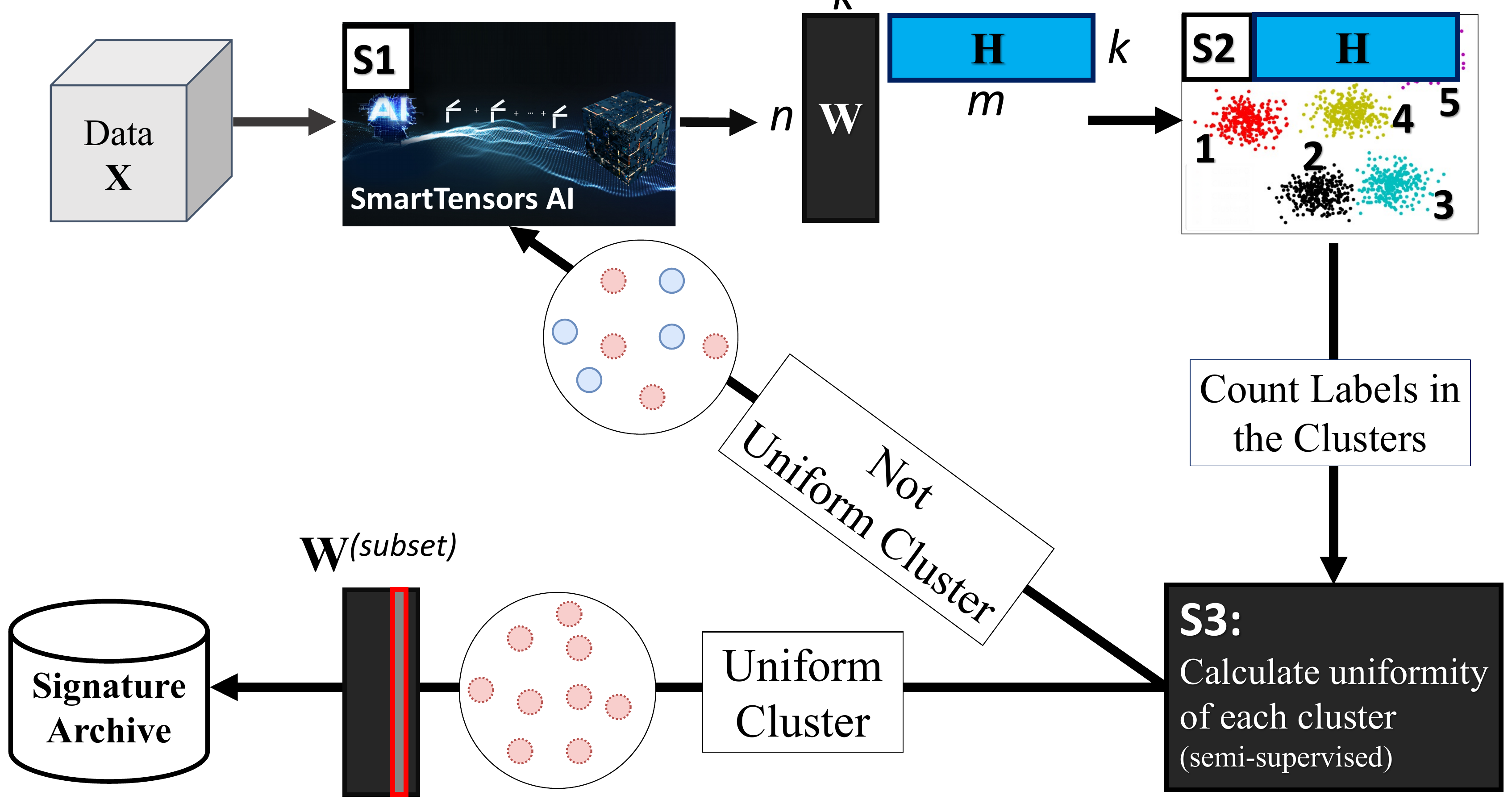}}
  \caption{Building the archive of latent signatures.}
      \label{fig:ALERTs}
  \vspace{-1.5em}
\end{figure}

\subsection{Building Signature Archive}

The overview of how the signature archive is built is shown in Figure \ref{fig:ALERTs}. MalwareDNA first applies NMF to the observational data $\mat{X}$ (\textbf{S1}). NMF is an unsupervised learning method based on a low-rank matrix decomposition~\cite{brunet2004metagenes}. NMF approximately represents an observed non-negative matrix, $\mat{X}\in \mathbb{R}_{+}^{n \times m}$, as a product of two (unknown) non-negative matrices, $\mat{W} \in \mathbb{R}_{+}^{n\times k}$ whose $k$ columns are the latent signatures each with $n$ features, and $\mat{H} \in \mathbb{R}_{+}^{k \times m}$ whose rows are the activities of each one of the $k$ signatures (latent features) in each $m$ samples, where usually $k\ll m, n$.  This approximation is performed via non-convex minimization constrained by the non-negativity of $\mat{W}$ and $\mat{H}$: min$||\mat[][ij]{X}-\sum^{k} _{s=1}\mat[][is]{W} \mat[][sj]{H}||_{F}$. 


The NMF minimization requires prior knowledge of the latent dimensionality $k$ for accurate data modeling, which is usually unavailable \cite{tan2012automatic}. Choosing too small a value of $k$ leads to a poor approximation of the observables in $\mat{X}$ (\textit{under-fitting}), while if $k$ is chosen to be too large, the extracted features also fit the noise in the data (\textit{over-fitting}). In this work, we use \textit{NMFk} that incorporates automatic model selection for estimating $k$ \cite{SmartTensors, nebgen2021neural}. NMFk integrates NMF-minimization with custom clustering and Silhouette statistics, and combines the accuracy of the minimization and robustness/stability of the NMF solutions, using a bootstrap procedure (i.e., generation of a random ensemble of perturbed matrices) is applied to estimate the number of latent features $k$. MalwareDNA uses a publicly available implementation of NMFk \cite{bhattarai2021pydnmfk}.


Next, we apply a custom H-clustering to assign each of the samples (the columns of $\mat{X}$) to one of the $k$ signature-clusters (\textbf{S2}). In each of these clusters, some of the samples may have different labels (non-uniformity) based on the confidence probability. We evaluate the uniformity of the samples in each cluster, determining whether all labels are the same (\textbf{S3}). When a uniform cluster is identified, we separate the samples of this cluster from the data, $\mat{X}$, and add the annotated (by the labels) cluster centroid, corresponding column of $\mat{W}$, to our archive of signatures. Otherwise, we continue with successive factorizations in a hierarchical manner to separate the mixed latent signatures as shown in Figure~\ref{fig:ALERTs}.

\subsection{Inference Using the Signature Archive}

During testing for real-time inference, we project each new sample onto the signature archive using Non-negative Least Squares Solver (NNLS). This allows us to perform real-time identification by representing each new sample as a combination of signatures recorded in the archive and estimating the accuracy, or similarity score, of this representation. We utilize the cosine similarity score of the NNLS projection of the new sample to the signatures in archive. We utilize the similarity scores, together with a threshold, $t$, to define the malware/benign-ware classification: When a signature possesses a similarity score above $t$, the labels of the signature will be determined as the classification result. Otherwise, when the similarity score is below $t$, it will be determined to be a novel malware family ($t=1.0$ in our experiments).

\section{Experiments}

To illustrate the capabilities of MalwareDNA, we randomly sample 1k benign-software and malware specimens from four families (ramnit, adposhel, emotet, and zusy) using a popular benchmark dataset, EMBER-2018 \cite{Anderson2018}. We select ramnit to represent a malware novel/unseen family. We use the static analysis features byte histogram and entropy, print table distribution, strings entropy, number of strings/exports/imports/sections, file size, and code size. 


\begin{figure}[htb]
\vspace{-1em}
  \centerline{\includegraphics[width=0.68\linewidth]{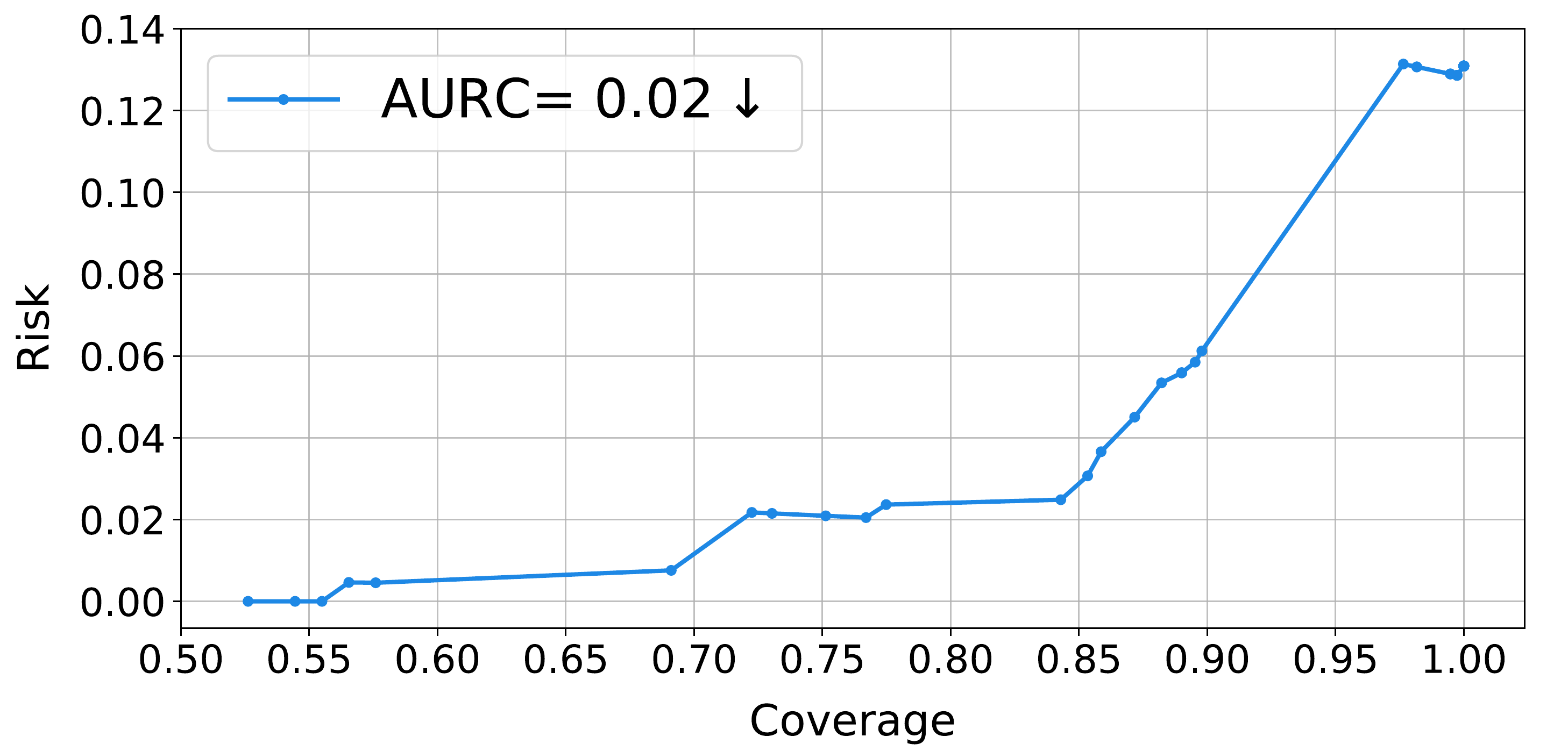}}
  \caption{Risk-Coverage (RC) curve when classifying malware families and the benign-software, together with the area under the RC (AURC).}
  \label{fig:aurc}
  \vspace{-0.5em}
\end{figure}

The performance of our method is reported with the Area Under the Curve of Risk-Coverage (AURC) \cite{ding2020revisiting} in Figure \ref{fig:aurc}. AURC models the trade-off between the coverage (the number of samples for which the non-rejecting predictions were made) and the risk which is measured with 0/1-loss. AURC score is reported between 0 and 1, and lower AURC is preferred over higher AURC. MalwareDNA achieve AURC of 0.02 when classifying the three malware families and benign samples. Our score indicates that we can achieve high coverage with minimal increase in the risk (false rejection predictions). 

\begin{table}[t!]
\caption{Performance of MalwareDNA compared to baselines. Rejection Seen provides the false rejection predictions for the samples that belongs to known classes. Rejection Novel is the true rejection predictions for the samples that belongs to a novel malware family. XGBoost+SelfTrain and LightGBM+SelfTrain achieve AURC score of 0.654 and 0.651.}
\label{table:baselines}
\resizebox{\columnwidth}{!}{%
\centering
\begin{tabular}{l|c|c|c|c|c}
\hline
\textbf{Model}     & \textbf{F1} & \textbf{Precision} & \textbf{Recall} & \textbf{Rejection Seen} & \textbf{Rejection Novel} \\ \hline
\textbf{MalwareDNA (ours)}  & \textbf{0.975}	     & \textbf{0.975}	          & \textbf{0.977}	        & 15.70 \%	              & \textbf{100.00 \%}             \\
XGBoost            & 0.416	     & 0.699	          & 0.510	        & NA                	  & NA             \\
LightGBM           & 0.297	     & 0.749	          & 0.338	        & NA	                  & NA          \\
XGBoost+SelfTrain  & 0.096	     & 0.258	          & 0.108	        & 4.34 \%	              &18.09 \%             \\ 
LightGBM+SelfTrain & 0.096	     & 0.078	          & 0.197	        & \textbf{2.89 \%}	              & 17.14 \%             \\ \hline
\end{tabular}
}
\vspace{-2em}
\end{table}

At 84.3\% coverege, MalwareDNA achieves an F1 score of 0.975 when classifying the malware families and the benign-software and 100\% true-rejection predictions for the chosen unseen family ramnit, which illustrates our method’s ability to identify novel malware families (Table \ref{table:baselines}). In Table 1, we also baseline our method against the state-of-the-art supervised malware classifiers XGBoost~\cite{chen2015xgboost} and LightGBM~\cite{ke2017lightgbm}. We further extend these baselines with the SelfTrain \cite{10.3115/981658.981684} algorithm to create semi-supervised models. We note that the previous work has used these models to report benchmarking against this dataset \cite{Anderson2018, marais2022malware}; however, we expose these models to a more challenging task of classifying malware families, separating them from benign samples, and detecting novel malware families all at the same time. Our baselines are tuned using Optuna \cite{akiba2019optuna} over 100 trials with 5-fold stratified shuffle cross-validation. Our benchmarking against the baseline models and the poor performance of these models, points out both the difficulty of the task, and MalwareDNA’ unique capability to both accurately detect malware, classify families, while simultaneously detect novel malware families.

\section{Conclusion}

In this paper, we introduced a new semi-supervised method that unifies three capabilities into a single framework: malware detection, malware family classification, and identification of novel malware families. Our preliminary results showcased the precise novel malware detection capability of our system while also outperforming state-of-the-art methods in a more difficult problem of solving all three inference tasks.

\section*{Acknowledgment}
This manuscript has been assigned LA-UR-23-25618. This research was funded by the LANL LDRD grant 20230753CR and the LANL Institutional Computing Program, supported by the U.S. Department of Energy National Nuclear Security Administration under Contract No. 89233218CNA000001.

\bibliographystyle{IEEEtran}
\bibliography{References}

\vspace{12pt}

\end{document}